\documentclass[manuscript]{aastex}

\begin{document}

\title{A Near-Infrared Stellar Census of Blue Compact Dwarf
Galaxies: The Wolf-Rayet Galaxy I~Zw~36\footnotemark[1]}

\author{Regina E. Schulte-Ladbeck}
\affil{University of Pittsburgh, Pittsburgh, PA 15260, USA}
\email{rsl@phyast.pitt.edu}
\author{Ulrich Hopp}
\affil{Universit\"{a}tssternwarte M\"{u}nchen, M\"{u}nchen, FRG} 
\email{hopp@usm.uni-muenchen.de}
\author{Laura Greggio}
\affil{Osservatorio Astronomico di Bologna, Bologna, Italy, and
Universit\"{a}tssternwarte M\"{u}nchen, M\"{u}nchen, FRG}
\email{greggio@usm.uni-muenchen.de}
\author{Mary M. Crone}
\affil{Skidmore College, Saratoga Springs, NY 12866, USA}
\email{mcrone@skidmore.edu}
\author{Igor O. Drozdovsky}
\affil{University of Pittsburgh, Pittsburgh, PA 15260, USA, and University of
St.~Petersburg, St.~Petersburg, Russia}
\email{dio@phyast.pitt.edu}

\footnotetext[1]{Based on observations made with the NASA/ESA Hubble 
Space Telescope obtained from the 
Space Telescope Science Institute, which is operated by the Association of 
Universities for Research in Astronomy, Inc., under NASA contract NAS 5-26555.}

\begin{abstract}

We report the results of near-IR imaging in J and H, of I~Zw~36 ($\approx$Z$_{\odot}$/14)
with the Hubble Space Telescope. Whereas imaging with the pre-COSTAR Faint Object Camera 
(FOC) previously resolved hot and massive stars in the near-UV, the NICMOS data  
furnish a census of the cool, intermediate- and low-mass stars. 
There clearly was star formation in I~Zw~36 prior to the activity 
which earned it its Blue Compact Dwarf/Wolf-Rayet galaxy classification. 
The detection of luminous, asymptotic giant branch 
stars requires that stars formed vigorously several hundred Myr ago. 
The well-populated red giant branch indicates stars with ages of at least 
1-2~Gyr (and possibly older than 10~Gyr). 
We use the tip-of-the-red-giant-branch 
method to derive a distance of $\geq$5.8~Mpc.  
This is the third in a series of papers on near-IR---resolved
Blue Compact Dwarf galaxies. We notice that the color-magnitude 
diagrams of VII~Zw~403, Mrk~178, and I~Zw~36 do not exhibit 
the gaps expected from an episodic mode of star formation.
Using simulated color-magnitude diagrams we demonstrate for I~Zw~36 that
star formation did not stop for more than a few 10$^8$~yrs over the past 10$^9$~yrs, and 
discuss the implications of this result.

\end{abstract}


\section{Introduction}

Nearly all galaxies in the low-redshift Universe show clear signs
of maturity: colors and spectra which suggest an older population of
stars in addition to any recent star formation. There are certain nearby
galaxies, however, which have long been suspected of youth. Extreme Blue
Compact Dwarfs (BCD) like I~Zw~18 exhibit extremely low metal abundances,
very blue colors, and large H-I reservoirs, and could be genuine 
proto-galaxies experiencing their first star-forming event
at the present epoch (Sargent \& Searle 1970; 
Searle \& Sargent 1972; Searle, Sargent \& Bagnuolo 1973;
Izotov \& Thuan 1999). It has been argued that if BCDs instead are old galaxies, then
they can nevertheless not have experienced more than a few, short starbursts separated 
by long quiescent periods.
Chemical evolution models predict that interstellar abundances rise above  
observational values after only a few star-forming episodes 
(e.g., Kunth \& Sargent 1986; Marconi, Matteucci \& Tosi 1994). 
With star formation rates (SFR) between 0.01 and 10~M$_{\odot}$yr$^{-1}$ 
and typical H-I masses of a few times 10$^8$~M$_{\odot}$, the gas depletion 
time scales for BCDs range from between a few times 10$^7$ to 10$^{10}$ years (Thuan 1991). 
Infrequent, short bursts have thus become widely accepted as the paradigm for the
star formation mode of BCDs (e.g. Thuan et al. 1999). This interpretation 
only recently has come under re-consideration (Legrand et al. 2000).

In the 1980s, deep ground-based CCD imaging revealed that most ($>$90\%)
of BCDs harbor an elliptical, red, low surface brightness sheet of background
stars (Loose \& Thuan 1986; Kunth, Maurogordato \& Vigroux 1988; also
Papaderos et al. 1996; Telles, Melnick \& Terlevich 1997; 
Doublier et al. 2000). With the advent of HST, it has become possible to 
resolve the background sheets of the most nearby BCDs into red giant branch 
(RGB) and asymptotic giant branch (AGB) stars. This avoids the degeneracy of 
spatially integrated broad-band photometry, which cannot distinguish 
intermediate- or low-mass stars from the massive, red supergiants (RSG). 
This paper on I~Zw~36 (12+log(O/H)=7.77, Izotov \& Thuan 1999; $\approx$Z$_{\odot}$/14) 
is the third in a series of papers which address the near-IR photometry 
of resolved stars in BCDs. The color-magnitude diagrams (CMDs) of VII~Zw~403 
(12+log(O/H)=7.69, Izotov \& Thuan 1999; $\approx$Z$_{\odot}$/17)
and Mrk~178 (12+log(O/H)=7.95; Kobulnicky \& Skillman 1996; 
$\approx$Z$_{\odot}$/10) were discussed in
previous papers. Both optical and near-IR CMDs were used 
to derive the star formation history of VII~Zw~403 
(Schulte-Ladbeck, Crone \& Hopp 1998, hereafter SCH98;
Lynds et al. 1998;
Schulte-Ladbeck et al. 1999, hereafter SHCG99;
Schulte-Ladbeck et al. 1999, hereafter SHGC99),
while near-IR CMDs formed the basis of the stellar census of
Mrk~178 (Schulte-Ladbeck et al. 2000, hereafter SHGC00). In 
both galaxies, star formation was detected at times that pre-date the current
activity, and both are at least 1-2~Gyr old. The very nearby, newly identified BCD NGC~6789 
was resolved from the ground (Drozdovsky \& Tikhonov 2000); and we just detected
the red giant branch with HST's WFPC2 (Drozdovsky et al. 2001a). Meanwhile, we also resolved
red giants with WFPC2 in the BCD UGCA~290, at a distance of about 
6.7~Mpc on the line of 
sight to the Canes Venatici cloud (Crone et al. 2000). We note that this
BCD does not share the dominant, ``iE" morphology; rather, it is a case
of a background sheet that was largely hidden to ground-based photometry underneath
the star-forming regions. This adds another two 
BCDs for which an old stellar component has been directly detected; and the results
for UGCA~290 suggest that ``Baade's red sheet" occurs in an even larger fraction 
of BCDs (over 90 \%) than was previously assumed. 

I~Zw~36 = UGCA~281 = Mrk~209 = Haro~29 is an excellent candidate with which 
to test existing formation scenarios for BCDs. It is nearby, has a low
oxygen abundance and large H-I content,  
and has been proposed to be a galaxy currently undergoing
its very first starburst (Fanelli, O'Connell \& Thuan 1988). Its 
extinction-corrected, total apparent B magnitude from the RC3 is 14.94, which, using our
new distance modulus of 28.8, translates into an absolute magnitude of
-13.9.

A comprehensive, multifrequency study of I~Zw~36 was presented by 
Viallefond \& Thuan (1983). They found a core-halo structure in H-I gas, 
with the visible star-forming regions situated near that core, but
slightly shifted with respect to the peak in H-I column density. This appears
to be quite a common occurrence in BCDs. Van Zee, Skillman \& Salzer (1998)
observed five BCDs in H-I and noticed they exhibit strong central
concentrations in their neutral gas distributions, but with moderate depressions
in column density over the most intensely star-forming regions.
The H-I data of I~Zw~36 of Viallefond \& Thuan (1983) also yielded the 
distance which has henceforth been used in the literature;
with a Hubble constant of 75~km~s$^{-1}$~Mpc$^{-1}$, they determined
4.6~Mpc. The H-I mass is then about 4x10$^7$M$_{\odot}$ (at a distance
of 5.8~Mpc, the H-I flux listed in Table~1 of Viallefond \& Thuan
corresponds to an H-I mass of 6x10$^7$M$_{\odot}$), and
the virial mass about 5--7 times the H-I mass. The latter result has pointed to
the potential importance of dark matter in this BCD. 
The nebular lines in the optical spectrum
and the blue-UV continuum show that star formation has been active in recent times. 
Interestingly, I~Zw~36 has an interstellar CO detection despite of its rather low
O abundance (Tacconi \& Young 1987). Near-IR photometry
was interpreted by Viallefond \& Thuan to indicate the presence of giant stars, rather 
than a contribution to the integrated colors by RSGs from the present star formation event. 
Using stellar CO indices to identify RSGs in two BCDs, this interpretation of near-IR colors 
was later called into question by Campbell \& Terlevich (1984).

In their CCD survey of BCDs, Loose \& Thuan (1986) identified an extended,
elliptical background sheet underlying the compact, active star formation core of I~Zw~36
(see their Fig~1). They classify I~Zw~36 as an ``iE" BCD; this is the most
common morphological type and considered characteristic of the BCD phenomenon. 

I~Zw~36 was included in the IUE UV spectral synthesis study of BCD stellar populations by
Fanelli, O'Connell \& Thuan (1988), who found that its spectral energy
distribution was accounted for by young, hot, O and B stars, with no light
from stars cooler than about B6. Fanelli, O'Connell \& Thuan 
pointed to the unique nature of I~Zw~36: 
unlike any other BCD in their sample, it {\it could} be undergoing 
its first starburst at the present epoch.

This finding prompted a follow-up UV imaging study with the pre-Costar HST (Deharveng
et al. 1994). Using the FOC with a variety of UV and blue filters, Deharveng
et al. successfully resolved massive stars in the inner 11~square~arcsec of
I~Zw~36. They noticed that the integrated F175W-F430W color of the galaxy
is redder than expected from the current star formation. Ruling out contamination by RSGs, 
as no RSGs appeared on their CMDs, they suggested the presence of
an older, underlying stellar population in addition to the young and massive stars.

Papaderos et al. (1996) subsequently obtained deep, ground-based optical imaging of I~Zw~36.
They drew attention to a second star-forming region spatially separated from the prominent
one, and proposed that a previous episode of star formation took place here. In addition,
they found that the galaxy becomes redder with increasing distance from
the core. They suggested this underlying low-surface-brightness component 
represents an older stellar population formed prior to the present starburst.

Most recently, the brightest supergiants of I~Zw~36 were resolved 
in the optical with
ground-based imaging. Makarova, Karachentsev \& Georgiev (1997) derive
a distance of 5.7~Mpc for I~Zw~36; placing this galaxy more than 1~Mpc further out than 
was previously assumed. I~Zw~36 lies on the line of sight to the Canes Venatici
cloud of galaxies.

In this paper, we show that I~Zw~36 resolves into single stars in the near-IR with HST.
We find that our near-IR CMDs clearly reveal the presence of RSGs in this galaxy.
Therefore, where young and old stellar populations spatially overlap, integrated
colors cannot distinguish the nature of the stellar content (see also SHGC99). 
The near-IR CMDs exhibit resolved AGB and RGB stars, the descendents of intermediate-
and low-mass stars. We derive a minimum distance of approximately 5.8~Mpc, 
and a minimum age of 1-2~Gyr for this galaxy.

\section{Observations and reductions}

We observed I~Zw~36 on 1998 July 7 as part of our GO program
7859. The NIC2 camera, which has a field of view about 19~arcsec across, 
was centered at (J2000) R.A. 12:26:16 and Dec. 48:29:39.
Figure~1 shows the location of the NIC2 observation relative to
the Digitized Sky Survey (DSS) image of the galaxy. The NIC2 field
overlaps with the FOC field. (The NIC2 total field shown in Fig.~2 is slightly
larger than the single NIC2 field superposed in Fig.~1, because 
the camera was dithered from one exposure to the next.) We shall not 
discuss the data in the NIC1 field, whose location is also indicated on Fig.~1. 
The NIC1 field contains very few (less than ten) stars, primarily because 
the sensitivity of this camera is lower than that of NIC2, and of course also because
of its smaller size and location near the edge of the visible galaxy body. While future data 
mining might allow us to gain information on any potential high- and intermediate-mass
stars at this location within I~Zw~36, our present
study focuses only on the more interesting, NIC2 results. 

The data were obtained in the F110W and the F160W filters, which are similar to
the ground-based J and H bands. The data acquisition consisted of two sets
of three exposures in F110W, and one set of three in F160W. The first
set of F110W exposures used integration times of about 896~s each.  All other
exposures were all about 960~s in length. The telescope was dithered by 
1$\arcsec$ between exposures. The data were read out
in MULTIACCUM mode.

The data were re-reduced following the steps described in SHGC99, and again
in SHGC00. In brief, the exposures were cleaned of cosmic rays; a temperature-dependent
dark was applied; they were inspected for cosmic-ray persistence; and
the pedestal was removed. All exposures were used in the combined,
final images in F110W and F160W. The total 
integration times were thus about 5568~s in F110W, and 2880~s in F160W. 
 
Fig.~2a,b illustrates how well I~Zw~36 resolves into single stars with NIC2. 
Comparing Fig.~2 with Fig.~1 of Deharveng et al. (1994), we recognize the most prominent
H-II regions were captured by both the near-UV and the near-IR imagers. 

Single-star photometry was conducted using DAOPHOT (Stetson, Davis, \& Crabtree 1990). 
The conversion from instrumental magnitudes to absolute photometry 
followed the NICMOS Photometric Calibration CookBook, with the photometric 
zero-points in the Vegamag system as provided by the NICMOS team. 
This yielded two photometry lists, F110W and F160W, in
the HST Vegamag system. Figure~3 shows the error distributions for the photometry. 
The photometry reaches limiting magnitudes of about
26.5 in F110W, and 25.5 in F160W. We merged the two star lists requiring a positional 
source coincidence of better than 1.5~pixels or $\approx$~0\farcs1. There are 
712 objects with coincident positions in F110W and F160W.
 
Completeness tests were conducted using the DAOPHOT/ADDSTAR routine, by adding 
100 false stars in each image and for each magnitude interval in the data. 
For each magnitude bin, 10 such tests were carried out.
 An artificial star was
considered to be recovered if the difference between input and output
magnitude was less than 0.75. 
The percentage of recovered stars is shown in Fig.~4.
These tests indicate that the data are better than 80\% complete to 
a magnitude of about 24 in F110W and about 23.5 in F160W.

Figure~5 shows CMDs constructed from these data. Subsequent CMDs in the
ground-system (J, H) use the transformations which we derived in SHGC99,
based on 
the NICMOS team's information on standard-star observations from their
photometry WWW site. Figure~6 shows the CMDs transformed to ground-based J 
and H, on the same scale as those of Fig.~5 for comparison. 

Another group has recently published transformations from F110W, F160W,
to ground based J and H in the CIT/CTIO system (Stephens et al. 2000). 
This group draws on observations of red, solar-metallicity stars in
Baade's window. We compared their transformation to our transformation, and found
there was no significant difference which would affect our results. For example,
a ``typical" stellar color near the tip of the red giant branch is F110W-F160W=1.0
in the HST Vegamag system.
In our ground system, this corresponds to a J-H color of 0.79$\pm$0.15. In the ground system
of Stephens et al., it corresponds to 0.61$\pm$0.14.
This amounts to a systematic color difference of 0.18$\pm$0.21 between the two systems. 
We will continue to use our own
transformations. Our distance calibration, and the transformation
of stellar-evolution tracks used in the simulators, are based on them (see SHGC99). 
For the above color, the difference in H magnitudes introduced by the
different calibrations is 0.05$\pm$0.16. Thus the distance
determination is not affected within the errors by our choice of calibration.

We note that any tracks which we overplot on subsequent CMDs originate from tracks 
in the flight system, which were transformed using the same transformation as 
the data whenever comparisons in the ground system are carried out. Thus, 
our data and our tracks are on a consistent system.

\section{Results}

Because the field of the NIC2 camera is so small and because I~Zw~36 is
situated at high galactic latitude (68$^o$), we expect the contamination
of the CMDs by galactic foreground stars to be very small. 
Based on the data of Ratnatunga \& Bahcall (1985),
we determine a galactic foreground of 0.7 stars in the NIC2 field of view
(down to a limiting magnitude of V=26).
Thus, even if this estimate should for some reason be by as much as a 
factor of 10 too small, it is still straightforward to interpret the CMDs.
The CMDs contain over 700 stars that must belong almost exclusively to I~Zw~36. 

The foreground galactic extinction towards I~Zw~36 is negligible: A$_B$=0.000
from H-I maps (Burstein \& Heiles 1984); A$_B$=0.065, which corresponds to
A$_J$=0.013 and A$_H$=0.009 from IRAS maps (Schlegel, Finkbeiner \& Davis 1998, and see NED). 

Like that of many other star-forming dwarf galaxies (e.g. Mas-Hesse \& Kunth 1999), 
the internal extinction of I~Zw~36 comes out
to be different from spectroscopy and imaging.
Deharveng et al. (1994) comment on how their CMDs do not suggest any
substantial extinction in the UV, whereas the spectroscopic Balmer decrement used
by Viallefond \& Thuan (1983) indicates E(B-V)=0.3. They also note that the UV spectrum
of Fanelli et al. (1988) indicates a very low internal extinction. 
We assume that, as in other star-forming dwarfs, reddening is much larger
for the ionized gas than for the diffuse stellar population (Calzetti et al. 1997)
and that the internal extinction is negligibly small in the near-IR CMDs.
 
The CMDs of Figs.~5 and 6 are characterized by a strong red plume, and a comparatively
weak blue plume (spread out around J-H=0). We definitely detect stars that qualify 
as RSGs belonging to
I~Zw~36. The near-IR observations thus complement the near-UV---optical 
observations from which the presence of RSGs remained doubtful (Deharveng et al. 1994).
Most of the stars in the near-IR CMDs appear at faint
magnitudes at the bottom of the red plume.  
{\it This is the red tangle which contains the RGB}. Depending on
the specific star formation history, 
some combination of AGB stars and even core-He-burning stars can contribute
to this feature. The dashed line of Fig.~5 marks the magnitude at which we identify
the TRGB (see below). Above the TRGB, we expect to see AGB and RSG stars. 
There is indeed a considerable number of red stars above the RGB.
The blue plume can include main-sequence (MS), blue-loop (BL), and blue-supergiant
(BSG) stars. It is not well defined owing to the large measurement errors
for blue stars in the data.

\subsection{The distance}

\footnotetext[2]{Drozdovsky et al. (2001b) confirm this calibration using
a larger number of low-metallicity galaxies.}

We use the tip-of-the-red-giant-branch (TRGB) method to constrain the
distance to I~Zw~36. There are advantages and disadvantages to attempting
the TRGB method in the near-IR. Of advantage is the fact that RGB stars
are brighter in J and H than in I and V, the two colors conventionally
used in the TRGB method (Lee, Freedman \& Madore 1993). We constructed
a calibration for the TRGB method in J and H using VII~Zw~403 HST data,
 stellar-evolution models, and globular cluster
data (SHGC99)\footnotemark[2]. Of disadvantage is the fact that CMDs of star-forming
galaxies, and those that involve near-IR colors in particular, 
always show a large number of AGB stars. 
Our near-IR observations of VII~Zw~403 and Mrk~178 
revealed that AGB stars are a strong stellar component (SHGC00); the same is
true for I~Zw~36. Low- as well as intermediate-mass AGB stars can  
overlap in color and luminosity with RGB stars near the TRGB. This causes 
luminosity functions of stars along the red plume to exhibit more of a gradual 
variation rather than the desired step function that would furnish a well-defined, 
TRGB magnitude for the distance determination. Finally, of all our near-IR observations, 
the errors and completeness fractions are poorest for
I~Zw~36. This by itself suggests I~Zw~36 is the most distant of the three
galaxies.

In determining the distance to I~Zw~36, we start by inspecting the
luminosity functions. The luminosity functions of 356~stars
selected to have colors in the range 0.7$<$(F110W-F160W)$<$1.4
and binned in 0.1~mag intervals are shown in Fig.~7. We applied an
edge-detection, Sobel filter (cf. Sakai, Madore \& Freedman  1996). 
The central value of the luminosity functions was varied in steps of 0.05~mag to reduce
the dependence of the results upon the particular choice of bin center.
The luminosity functions were then smoothed using a sliding average of
three points, and a Sobel kernel, [-1,0,+1], was applied. 
The position of the TRGB is indicated by the highest peak that
falls shortward of the completeness limit, at m$_{F110W_0, TRGB}$~=~24.52 and
m$_{F160W_0, TRGB}$~=~23.46, respectively. Because we adopted a 0.1~mag binning
in the initial luminosity functions,
the formal error on these positions is $\pm$~0.05~mag. 

We derive the distance modulus of I~Zw~36 using the
calibration established for VII~Zw~403 in SHGC99:
M$_{F110W_0, TRGB}$ = --4.28~$\pm$0.10 $\pm$0.18  and
M$_{F160W_0, TRGB}$ = --5.43~$\pm$0.10 $\pm$0.18,
where the first error is the statistical error and is dominated by
how well we can determine the location of the TRGB in VII~Zw~403, and the
second is the systematic error primarily due to the RR~Lyrae distance
calibration of the TRGB (see SCH98). Application of the above calibration
assumes that the RGB stars of I~Zw~36 have the same metallicity as those
in  VII~Zw~403 ([Fe/H]=--1.92). The F110W-F160W color at the TRGB of 
I~Zw~36 is certainly consistent with assuming a low metallicity for the 
RGB. In what follows, the errors quoted are those
that refer only to our statistical errors. 
Applying this calibration directly yields the following distance moduli 
for I~Zw~36 using the TRGB in F110W and F160W, respectively: 
28.8($\pm$0.1) and 28.9($\pm$0.1), 
corresponding to distances of 5.8($\pm$0.3)~Mpc and 6.0($\pm$0.3)~Mpc.    
 
In SHGC99, we derived a more general calibration over a wider range of metallicities 
by comparing the H-band magnitude of the TRGB of VII~Zw~403 with that of GCs and
stellar evolution models. The TRGB in H is constant as a function of metallicity 
for a wide range of low metallicities: M$_{H_0, TRGB}$ = --5.5$\pm$0.1 
for --2.3$<$[Fe/H]$<$--1.5.
The transformed H magnitude at the TRGB is m$_{H_0, TRGB}$ = 23.4$\pm$0.1
(where the error is our transformation uncertainty).
This suggests an H-band based distance modulus of 28.9($\pm$0.2), corresponding to 
a distance of 5.9($\pm$0.4)~Mpc.

To summarize, the distances derived using F110W, F160W and H agree very well.
We shall therefore assume a distance modulus of 28.8 for our minimum
distance to I~Zw~36. We inspected Fig.~5 to double-check that the inferred ``jumps"
of stellar numbers in the luminosity functions, on which we base the
distance estimate, coincide with where
we would place the TRGB on the CMDs. As was the case in Mrk~178 (see
SHGC00), the top of the red tangle which includes the RGB is quite wide, 
and it is markedly tilted in the [(F110-F160), F160]
CMD (see Fig~5). We could place the TRGB in F160W at 0.2~mag fainter, and it would still lie
within the rather broad rise of the F160W luminosity function shown in Fig.~7.
We also notice that we find the TRGB at magnitudes
where the completeness of the data has already begun to drop to around 80\%
(see Fig.~4 and section 2). Our artificial star experiments show that
the recovered stars are systematically brighter than the
input stars at these magnitudes. This is due to blending, which has the 
effect of smearing out the discontinuity
of the number counts at the TRGB, artificially brightening its
location. Therefore, the real uncertainty of the tip detection
could be larger than the formal error we assumed so far.
While it is unlikely that we find the TRGB at a location that is
too faint, the above considerations indicate it is
probable that we find it too bright. Therefore, a more
realistic error on the distance modulus is not symmetrical.
We adopt 28.8$^{+0.2}_{-0.1}$ as our best estimate. 
As stated above, the systematic errors amount to an additional uncertainty 
of $\pm$0.18. The distance of 5.8~Mpc used throughout this
paper represents the minimum distance to I~Zw~36.

As we do not have a metallicity calibration
for the RGB in the near-IR, our method 
gives the smallest distance which is consistent with the data.
If, however, the RGB population of 
I~Zw~36 is more metal-rich than assumed above, then the TRGB indicates
a larger distance modulus. This is due to the metallicity dependence
of M$_H$ (as illustrated in Fig.~11 of SHGC99). The high-metallicity
distance modulus of I~Zw~36 (corresponding to the TRGB on the Z=0.004
tracks) is 29.5; this is our maximum distance. 
Therefore, an additional uncertainty of +0.7 in the distance modulus 
comes about due to unknown RGB metallicity.

Our minimum distance agrees very well with the distance determined from the 
brightest stars method, 5.7~Mpc, by Makarova, Karachentsev \& Georgiev (1997).  
In the subsequent sections of this paper, we compare the I~Zw~36 data with theoretical tracks
of stellar evolution and with simulated CMDs. We point out that any ``high-metallicity" 
assumption for the RGB corresponds to a ``long" distance.  From our previous, optical work on 
VII~Zw~403 and UGCA~290 we have a slight preference for the ``low-metallicity" assumption.
The assumption of a low-metallicity RGB is  supported 
by the empirical fact that other dwarf galaxies with low oxygen abundances in their 
H-II regions also tend to have low iron abundances of their old stellar populations
(cf. Fig.~1 of Kunth \& \"{O}stlin 2000). 
However, without better knowledge of the metallicity
of the RGB stars in I~Zw~36, we emphasize that the ``low-metallicity" or ``short" 
distance used here must be
regarded as a lower limit to the true distance of I~Zw~36.

\subsection{The stellar content}

Which of the features observed on the CMDs should be considered real,
and which artifacts of the errors?
In Figure~8, we display [(J-H)$_0$, H$_0$] CMDs of I~Zw~36 with three 
values of the DAOPHOT errors applied as a selection criterion.
These may be compared with Fig.~6. When both the J and H errors are smaller than
0.1~mag, the dominant contribution is a bright, red feature attributable to RSGs and
luminous AGB stars. We see only very few BSGs and/or massive MS stars. 
When the J and H errors are smaller than 0.15~mag, 
both the blue and red
plumes of the CMD become better populated. We also pick up some RGB stars; however,
the TRGB is not yet obvious (although there is a ``break" in the data
of the red plume below H$_0$ of about 23~mag).  
If the magnitude errors are further increased to 0.2~mag, 
all the salient features of the CMD are apparent. 
At this error cut, the
thermally-pulsing, or TP-AGB stars, defined as having (J-H)$_0$ 
colors redder than 1~mag (beyond the end of our tracks, see Fig.~9),
become visible in larger numbers. The
CMD of I~Zw~36 indeed contains quite a number stars of such red color, but many have
high photometric errors.

In our discussion, we employ several sets of tracks. The first set is based
on the ``old" Padova tracks which are presently incorporated in our CMD simulator.
Briefly, these tracks are based on the Padova library 
(e.g. Fagotto et al. 1994, Bertelli et al. 1994) and the stellar 
atmospheres of Bessell, Castelli \& Plez (1998).
A ``new" database of Padova tracks for low- and intermediate-mass stars
recently became available (Girardi et al. 2000), and we considered these as well. 
We note that the changes were not so substantial that we
felt compelled to replace the old tracks in our simulator.
In Figure~9, we overplot onto the [(J-H)$_0$, H$_0$] CMD  
stellar evolutionary tracks from the ``old" database. 
We show two sets of tracks with metallicities of
Z=0.0004 and Z=0.004 (Z$_\odot$/50 and Z$_\odot$/5, respectively).
The present-day ionized gas abundance of I~Zw~36 is bracketed
by these two. Tracks with Z=0.001 (the closest match to the oxygen abundance
in the database)
fall inbetween the tracks of Z=0.0004 and Z=0.004, and so the tracks shown 
in Fig.~9 are a reasonable approximation for any stellar population
which we might expect to find. The tracks do not extend to the extremely 
red colors observed for some of the asymptotic
giants. This is due to the truncation at the first thermal pulse 
of the electronically available Padova tracks.

There are many indicators of present-day star-forming activity in
I~Zw~36. The blue plume of the CMD of Fig.~9 is consistent with
massive MS stars and BSGs. The MS turnoff of the 30~M$_\odot$ track 
is located well within the blue plume. Deharveng et al. (1994)  
suggest an age of less than 12~Myr for the present star-forming
event based on the stellar content observed in the FOC-UV images. 
Star formation in I~Zw~36 
must have been active within the last few Myr because it contains 
spectroscopically detected Wolf-Rayet stars
(Izotov, Thuan \& Lipovetsky 1997). The UV spectrum indicates
massive stars via their stellar wind lines, and the H-II regions 
further support the presence 
of ionizing, young and massive stars with ages of up to 10~Myr. 

The red plume is the dominant feature of the I~Zw~36 CMD in the near-IR. 
It can be split into regions  
above and below the TRGB. The luminous portion of the red plume can in 
principle be composed of RSG stars and AGB stars. 
If we use the oxygen abundances in the ionized gas as an indication of the
present-day metallicity and compare the upper CMD with the Z=0.001 tracks, 
then clearly some of the stars in the upper red plume must be interpreted as RSGs. 
These stars have ages of a few ten Myr. RSGs
were considered to be ``missing" from the Deharveng et al. (1994) optical---near-UV 
CMDs, where the small number of red stars detected had large photometric 
errors and/or was accounted for by red leak and Galactic foreground contamination.

Comparing the two sets of tracks shown in Fig.~9 with the data illustrates a possible
age-metallicity degeneracy for the luminous, red stars:  we could 
interpret the few very high-luminosity stars 
at $M_H$ of about -10 to -9 and colors (J-H)$_0$~$\approx$~0.9 and 1.5 either as 
``high-metallicity"
 RSGs and thus, early descendants of massive stars, or as the 
TP-AGB phase of intermediate-mass stars of a ``low-metallicity". 
In the latter case these
stars would lie on extensions of intermediate-mass tracks beyond the
first thermal pulse. However, we do not know whether 
a low-metallicity TP-AGB star would ever become luminous and red enough to 
match the most luminous and red stars in our data. As Fig.~12 suggests,
there is one star classed AGB in I~Zw~36 which is somewhat brighter than those
classed AGB stars in VII~Zw~403 and Mrk~178 (cf. also, Fig.~11 of SHCG00), and
we should not attach too much weight  to this singular
data point. On the other hand, observations of AGB stars
in the LMC suggest stars with M$_H$$\approx$-9.5 are observed (Trams
et al. 1999). If there are any AGB stars in the upper red plume, then their
ages are over a hundred Myr.   

We can more clearly separate the AGB from the RSG stars observationally when 
they are redder than (J-H)$_0$~$\approx$~1.0, which is also in the regime of the 
thermally pulsing, or TP-AGB phase. Indeed we see in Figs.~8 and 9 quite a few stars
which are bright and red, (J-H)$_0$~$\approx$~1.5.
TP-AGB stars in principle probe ages of a few hundred~Myr to a few Gyr. 

Stars below the TRGB are interpreted to consist in part of low-mass, red giants.
Only low-metallicity sets of tracks can correctly describe the
blue color of the red tangle; the Z=0.004 tracks are too red.
In other words, while the Z=0.004 tracks can provide a fairly good
description of the luminous stellar content of I~Zw~36, they appear
too red for the bulk of the RGB stars.
This is similar to what we found for Mrk~178. 

The comparison of the data with the tracks leads to the conclusion that
we cannot describe the evolving stellar content with theoretical models of a single
metallicity. From Fig.~9 it
can be seen that the Z=0.004 tracks have too red a RGB, while the Z=0.0004 
have too blue an AGB location with
respect to the data. The Z=0.001 low-mass tracks can match the color of the
RGB, but, like the Z=0.0004 set, they also remain too blue with
respect to the AGB stars. Thus, we could not find a single value for the
metallicity which would simultaneously reproduce the colors of the old, the 
intermediate-age, and the young stars. The most obvious solution 
to achieve a better match is to allow some metallicity evolution with time 
in I~Zw~36.  

\subsection{The morphology}

The NIC2 camera was located on the highest-surface-brightness area of
I~Zw~36 as seen from the ground. Fig.~2 shows that the strongest
star formation is centrally located on the chip. The near-IR-resolved stars 
and the gas are found to overlap spatially with the elongated plume of UV-detected 
massive stars and gas which crosses the FOC chip (see Fig.~1 of Deharveng et al. 1994). 

Figure~10 shows the NIC2 coordinates of the stars in the
blue plume of our CMD, selected to have (J-H)$_0$$<$0.2.
Blue stars are resolved near the center and toward the lower right of NIC2, 
coinciding with the brightest H-II regions and the massive stars resolved in the UV.
Our CMD, in addition to confirming the presence of young and massive stars,
also exhibits the red tangle, that feature which contains intermediate- and low-mass
stars and potentially harbors the truly ancient population of this galaxy. 
Fig.~10 illustrates how stars in the red tangle, which were selected to have 
(J-H)$_0$$>$0.5 and H$_0$$>$23.3, display a fairly uniform 
spatial distribution across the face of the galaxy (disregarding the ``holes"
in the center which are due to the high level of crowding in the star formation regions). 

Surface-brightness profiles of I~Zw~36 throughout and beyond the
area of the galaxy encompassed by our NIC2 observations were 
investigated by Papaderos et al. (1996). They confirm I~Zw~36 as 
a type ``iE" BCD in the classification scheme of Loose \& Thuan (1995). In iEs, 
the faint outer isophotes are elliptical and the bright star formation regions are found
in the vicinity of the center, although not necessarily at the
exact center. 
As can be seen from Fig.~3a of Papaderos et al., the outer isophotes 
of I~Zw~36 are quite extended. Papaderos et al. give surface-brightness
profiles out to a distance of 40$\arcsec$ from the center of the galaxy.
Recall that the FOV of the NIC2 is about 19~arcsec across. The B-R color profile 
shows that I~Zw~36 becomes redder with increasing distance from the center. As 
illustrated by Fig.~3e of Papaderos et al., the color first rises steeply outward 
from the core until a B-R color of about 1.0$\pm$0.1~mag is achieved at a distance of
about 20$\arcsec$. At larger distances, the color remains constant at this level, 
until the profile is lost in the noise. The large color gradient in the inner 
20$\arcsec$ of the color profile has an obvious correspondence with the distribution 
of the resolved stars detected on NIC2 --- luminous, blue stars dominate 
in the center, while faint red stars are seen everywhere.
 
We may age-date the stellar content at large distances (where the color
gradient has flattened out), by comparing the B-R color to 
empirical cluster colors from Schmidt, Alloin \& Bica (1995). 
Such a comparison of course yields a luminosity averaged age.
A B-R color of about 1.0$\pm$0.1~mag, 
lies between the cluster templates with age 0.7-2~Gyr (B-R=0.91),
and age 2-7~Gyr (B-R=1.17), or the dwarf Elliptical/
dwarf Spheroidal (dES) galaxy template (B-R=1.17) of Schmidt, Alloin \& Bica.
As an alternative, we also compared the B-R color to theoretical, spectral evolution
synthesis models. The models by Bruzual \& Charlot (2000) indicate that,
for a single stellar population ages from about 1 to 9~Gyr are possible, 
with lower ages corresponding
to our high metallicity, Z=0.004, and higher ages corresponding to our
preferred, low metallicity, Z=0.0004, assumption. Thus, both comparisons
agree with our inference from the CMDs that stars with ages of at least 1-2~Gyr
and probably much older, are present in I~Zw~36.
The faint red low surface brightness component of I~Zw~36 can be traced 
to at least 40$\arcsec$ from the center. 
At a minimum distance of 5.8~Mpc, the physical diameter of the low-surface
brightness sheet of potentially ancient stars
in I~Zw~36 is thus at least 2.3~Kpc.

We previously illustrated, using the CMD of VII~Zw~403, how small fractions of luminous
young stars dominate integrated photometry even in the near-IR (see Table~2 of SHGC99). 
In H-band photometry of VII~Zw~403, for instance, luminous RSGs (and AGB stars) amount 
to about 63\% of the 
integrated light on the NIC2 chip. Origlia \& Oliva (1998) make use of 
this effect
to estimate, for a sample of 35 BCDs, the mass involved in the burst relative to that 
in the old stellar component. They find that 
in most BCDs, including I~Zw~36, the burst mass is much smaller than the mass of the
older, underlying stellar population.  
These results confirm and expand on the earlier findings of Papaderos et al. (1996).
Accordingly, the total luminosity of I~Zw~36 is divided into about equal parts between 
the burst luminosity and that of the old component of the host galaxy (Papaderos et al. 1996).
For any reasonable mass-to-light ratios for the two components, the old component
must dominate the total stellar mass. 
We also found such a result using simulations of the near-IR CMD 
of Mrk~178 (SHGC00). Therefore, the stellar mass of most BCDs is apparently 
locked up in low-mass stars and 
not in the few massive stars which dominate the light. This raises the mass 
fraction of BCDs which must be attributed to baryons, although it is
difficult to tell by how much (an attempt and a discussion of the pitfalls may be found in
SHGC00).

We find from the comparison of the NIC2 data and the surface-brightness profiles,
that I~Zw~36 has a core-halo structure. The central core is defined by
the star formation regions; here the luminous stars dominate integrated photometry. 
Areas of the galaxy that fall outside of the NIC2 FOV, are red and of low-surface 
brightness. We interpret this to indicate that where the young, massive stars are absent, 
the integrated light comes from the intermediate- and/or low-mass stellar content 
of I~Zw~36. Notice that prior to resolving stars in the background sheet with HST,
it was not possible to be certain that the red color of the background sheet was due to 
RGB and AGB stars rather than to RSG stars. Since copious bright RGB stars first appear at
an age of about 1~Gyr in populations of a wide range of metallicities (e.g., 
Sweigart, Greggio \& Renzini 1990), we quote this as our low age limit. 
The age estimate derived from the integrated B-R color at large distance is consistent with 
this minimum age. Both CMDs and integrated colors, in principle do allow larger ages for the
oldest stars. However, age-metallicity degeneracy prevents us from 
claiming a specific age for the oldest stellar population in I~Zw~36.

\subsection {The star formation history}

\subsubsection {The present-day SFR}

A quantitative SFR based on the IUE spectrum of I~Zw~36 was
derived by Fanelli, O'Connell \& Thuan (1988). The UV 
measures the fluxes of massive, newly-formed stars close 
to the maximum of their spectral energy distributions 
and is in principle one of the best gauges of SFR.
In practice, the main drawbacks of UV-derived SFRs are their sensitivity 
to extinction and on the shape of the IMF.
Fanelli, O'Connell \& Thuan combined
two spectra of I~Zw~36 taken through IUE's 10" x 20" aperture, which were
slightly offset from one another, but basically centered on the
brightest regions. They derive a SFR of 2$\times$10$^{-3}$~M$_\odot$~yr$^{-1}$
from modeling the UV spectrum with a reasonably small extinction
(i.e., not that which is indicated by the Balmer decrement).
This SFR is for stars with masses between 2 and 100~M$_\odot$,
assumes an IMF index of 2.8, and was scaled by them from their
empirical measure involving only stars with masses above 10~M$_\odot$.
If we adopt instead the Salpeter IMF (2.35) with limits 
of 0.1 and 100~M$_\odot$ (as is commonly used in dwarf galaxy research)
then we find a rate of 1.7$\times$10$^{-2}$~M$_\odot$~yr$^{-1}$.
This rate is based on a distance of 4.6~Mpc; if we
rescale to our minimum distance of 5.8~Mpc, then the SFR becomes
2.7$\times$10$^{-2}$~M$_\odot$~yr$^{-1}$. Fanelli, O'Connell \& Thuan
also provide an upper limit of 10~Myr on the duration of the
star-forming event seen in I~Zw~36.

We derive a quantitative SFR directly from the 
massive stellar content observed on our CMD below. At our minimum distance
the maximum rate which is consistent with the number of blue stars seen in
the NIC2 field of view (about 19 square arcsec) is 
2.5($\pm$0.4)$\times$10$^{-2}$~M$_\odot$~yr$^{-1}$. 
A principal concern with this CMD-based SFR is that it is constrained by 
a very small number of the most massive stars, only. 
The detection limits we achieve correspond to the
main-sequence turnoff of a 15~M$_\odot$ star; and
we are extrapolating rather a long way along the Salpeter IMF.

Other popular gauges of the SFR count the population of
massive stars indirectly. For example, the nebular emission from
gas ionized by massive stars is a widely used measure. 
In what follows, we determine the SFR using H$_{\alpha}$ luminosity
converted into SFR with
the equation introduced by Hunter \& Gallagher (1986).
We correct only for foreground extinction (using the galactic extinction from
Schlegel, Finkenbeiner \& Davis (1998), with the Cardelli, Clayton
\& Mathis (1989) extinction law), realizing that
this yields a lower limit to the actual SFR. The spectrophotometric
H$_{\alpha}$ flux of Viallefond \& Thuan (1983) determined in a 6\farcs1 circular aperture 
leads to a SFR of 1.9$\times$10$^{-2}$~M$_\odot$~yr$^{-1}$ for I~Zw~36. 
As this flux was derived in an aperture which is small
compared to the size of the galaxy, some fraction of the total flux from
H-II regions as well as diffuse, ionized gas, might have been missed. 
A new value for the H$_{\alpha}$ flux was
recently measured by Hunter (2001). Using imaging observations, she
obtains 1.29$\times$10$^{-12}$ erg~cm$^{-2}$~s$^{-1}$. 
This converts to a SFR of
3.8$\times$10$^{-2}$~M$_\odot$~yr$^{-1}$. As discussed above, 
it is not clear how dust is distributed internally within
I~Zw~36. Therefore, we prefer not to correct the H$_{\alpha}$ flux 
with the large internal extinction derived from the Balmer decrement. 
The uncertainty in how to apply an internal extinction correction is
always a major limiting factor to the accuracy of H$_{\alpha}$-derived
SFRs. 

The SFR can also be determined assuming that the radio continuum
emission arises from synchrotron radiation produced by relativistic
electrons that are accelerated when massive stars explode as supernovae 
(Condon 1992, but see Viallefond \& Thuan 1983). Radio fluxes
suffer much less extinction than UV and optical ones. I~Zw~36 was included in the NRVO/VLA
Sky Survey (NVSS) and its total 1.4~GHz flux is catalogued,
4.5~mJy (Condon et al. 1998). We find a SFR of 2.2$\times$10$^{-2}$~M$_\odot$~yr$^{-1}$ 
(with an extension of Condon's 1992 low-mass cut-off from 5~M$_\odot$ to 0.1~M$_\odot$
along the Salpeter IMF). The SFR from the radio continuum comes out to
be smaller than the lower limit on the SFR derive from H$_{\alpha}$ emission.
This is a known effect. Cram et al. (1998) show that H$_{\alpha}$ luminosities
give systematically higher SFRs (by as much as an order of magnitude)
than 1.4~GHz luminosities for galaxies
with small SFRs (smaller than about 10$^{-1}$~M$_\odot$~yr$^{-1}$). The
reason for this discrepancy has not yet been resolved.

We see that any and all methods available to determine the present-day
SFR of I~Zw~36 have some kind of caveat. We adopt a SFR of
2.5$\times$10$^{-2}$~M$_\odot$~yr$^{-1}$ as our best estimate.

\subsubsection {Comparing the near-UV and near-IR CMDs of I~Zw~36}
 
The near-UV---optical CMDs of I~Zw~36 by Deharveng et al. (1994) lack evolved red supergiants.
This cannot imply a discontinuous SFH, since the near-IR CMDs show such a copious
population of luminous red stars. To explain the absence of red supergiants on the 
near-UV---optical CMDs, we consider two observational biases: 
color bias and spatial bias.

To illustrate color bias, we reproduce in Figure~11 from Table~1 of Deharveng et al. (1994)
the near-UV---optical [m342-m430, m430] CMD of I~Zw~36, and compare it with our 
[(J-H)$_0$, H$_0$] CMD. The near-UV---optical CMD of I~Zw~36 approximates U-B. It
exhibits a strong blue plume from
the present generation of massive stars. Deharveng et al.
age-date the young generation of stars at $\leq$12~Myr (but do
not give a SFR). A few stars are observed in the red plume. 
Deharveng et al. discuss these stars explicitly. They find that most of them 
are not real owing to red leak, large error bars, and the possibility of
foreground contamination. They conclude that no, or very few, RSGs exist
in I~Zw~36. This supports their burst age.
However, since the integrated color of I~Zw~36 is redder
than would result from just the ongoing star formation,
Deharveng et al. propose the existence of an underlying background sheet of 
much older stars. This is consistent with a SFH
that had a long quiescent period prior to the current event. 
By contrast, the same region of I~Zw~36 observed in the near-IR reveals a strong component 
of luminous, red stars. These are in part the ``missing" RSGs. The detection
of copious AGB stars in particular extends the star formation of I~Zw~36 to times that pre-date
the event seen in the near-UV---optical CMD. These stars
suggest a more or less continuous history of star formation over the last Gyr. 
In essence, the near-UV---optical CMD is best used to derive the stellar content 
for the past 10~Myr, while the near-IR CMD yields a census for the past several 100~Myr.  

Spatial biasing could also have affected the Deharveng et al. results. As we now know,
the stellar content of BCDs varies as a function of position. 
Fig.~1 illustrates that the NIC2 and FOC pointed to roughly the same region
in the star-forming core of I~Zw~36.
However, massive, young stars completely filled the small FOV of the FOC, possibly
masking faint red stars due to crowding (compare Fig.~10).
Their images also exhibit a large contribution by ionized gas. This possibly made stellar 
detections difficult for DAOPHOT, since it expects a flat background and does not
perform well in the presence of a spatially nonuniform background 
(Stetson, Davis, \& Crabtree 1990).

In summary, this section serves as a note of caution that SFHs derived from CMDs
may be affected by observational biases which, though they can in part be addressed 
with simulated CMDs, may prevent us from accurately recovering in all cases the actual 
SFH of a galaxy.
It also serves to show that any gaps on CMDs need to be carefully scrutinized
for observational effects before concluding a discontinuous SFH. 

\subsubsection {Intercomparison of the near-IR CMDs of BCDs}

Near-IR photometry to deep limiting magnitudes is still a new
technique for stellar population studies.
In this section, we compare the near-IR CMD of I~Zw~36 with that
of two nearby BCDs, VII~Zw~403 and Mrk~178, and with that of I~Zw~18. 

We previously resolved VII~Zw~403 and Mrk~178 
in the near-IR to such deep limiting magnitudes that we were able to 
classify their stellar content down to magnitudes below the TRGB. This classification scheme
was originally developed by comparing the relatively well-understood
V, I CMD of VII Zw~403 with the locations of the same stars on its
J, H CMD (SHGC99).  We used this classification to investigate the 
J, H CMD of Mrk~178 in SHGC00. 
In Fig.~12, we apply the classification to I~Zw~36. We note that
these three galaxies are all relatively nearby and that the data have similar
crowding and blending properties. (In Fig.~12, the red tangle of VII~Zw~403
extends to blue colors, and the blue plume extends into the red tangle, at the faintest
magnitude levels. This is due to the fact that the near-IR data have different
errors than the optical data from which the classification was derived. 
While we place the blue vs. red plume dividing line at 0.5~mag in Fig.~12,
we plotted stars with (J-H)$_0$$<$0.2 only as blue plume
of I~Zw~36 in Fig.~10, to safely split off the blue plume from the red tangle.)

In Fig.~12, we color code all stars with (J-H)$_0$ color smaller
than 0.5 in blue. The purpose is to illustrate that we consider these stars 
to be part of the blue plume. Stars with colors 0.5$<$(J-H)$_0$$<$0.85 and H$_0$ above the
TRGB are shown as RSGs in magenta. Stars with colors $>$0.85 are 
AGB stars and shown in black. This coding combines the early, or E-AGB, and the
TP-AGB phases. Note that stars in the TP-AGB phase are expected to have 
(J-H)$_0$$>$1. The remaining stars are color-coded in red and considered to belong to
the red tangle. 

Comparing the stellar content of VII~Zw~403, Mrk~178, and I~Zw~36 in this way, we find
a great deal of similarity. All the major phases of stellar evolution are
represented on the CMD. In SHGC00 we presented a set of exemplary 
CMD simulations in the near-IR
for Mrk~178. (We note that in order to avoid color biasing, the present-day SFR was
derived from the H$_{\alpha}$ luminosity as well as from simulated CMDs.)
Since the near-IR CMD of I~Zw~36 exhibits the same morphology
as that of Mrk~178, the star formation histories are similar. In particular, the strong
AGB stellar component suggests that star formation was very active in the recent past, i.e.,
a few hundred Myr ago. A similar result, enhanced
star formation in the interval 600--800~Myr, was
gleaned from simulations of the optical CMD of VII~Zw~403 (Lynds et al. 1998).

The famous I~Zw~18, a galaxy still hotly contested to be forming
its first stars in the present burst, was observed with HST/NIC2 by \"{O}stlin (2000). 
Its near-IR CMD is difficult to compare 
to that of I~Zw~36. The CMD contains only about 70 stars, 
compared to over 700 stars in I~Zw~36. Therefore, the
stochastic properties of the two datasets are quite different. I~Zw~18 is at about
twice the distance of I~Zw~36 (\"{O}stlin adopts 12.6~Mpc). Owing to the
larger distance of I~Zw~18, its NIC2 data have different blending and crowding
properties than those of I~Zw~36.  \"{O}stlin infers a more or less continous star formation
over the last 1~Gyr; a SFR peak at 14~Myr is identified with the current starburst. 
The optical, HST/PC2 CMD of I~Zw~18 was previously modeled by Aloisi, Tosi \& Greggio (1999), 
who adopted a distance of 10~Mpc. From the comparison of observed and synthetic
luminosity functions they find a single starburst with a duration of order 10~Myr 
is definitely ruled out. However, a single episode can reproduce the data if it
started more than 200~Myr ago and either continues today or stopped
5~Myr ago. A two-event 
scenario is also possible and preferred by Aloisi, Tosi \& Greggio, 
with a first episode from 1~Gyr to 30~Myr ago at constant SFR, and a second one, 
7---10 times stronger, from 20 to 15~Myr ago. In spite of
the ample amount of HST time devoted to observations of I~Zw~18, the conclusions about
its recent SFH are still ambigous and await, in particular,
a distance determination based on a stellar distance indicator. 

To summarize, we find that the SFH of I~Zw~36 can be well traced over a wide range of ages 
with the aid of near-IR CMDs. There do not appear to have been any distinct, strong bursts 
and long quiescent periods. This conclusion is further supported by
our models of the stellar distribution on the CMD, presented below.

\subsubsection {The mode of star formation}

We now turn to the question of what exactly is  
the star formation mode of nearby BCDs. 
Stars are distributed across the major areas of the CMDs of
VII~Zw~403, Mrk~178, and I~Zw~36. They represent all the stellar masses and
major stellar evolutionary phases which are above the detection limits. 
There are no obvious gaps in the distribution
of stars. Gaps on the CMDs are expected for BCDs; their
star formation is commonly viewed as episodic. For instance, Mas-Hesse \& Kunth (1999)
find a very short mean duration of 3.5~Myr for the starbursts in a sample of BCDs and
dwarf Irregular (dIrr) galaxies, while Thuan et al. (1999)
envision quiescent periods of 2-3~Gyrs between the active starburst phases of BCDs. 
This mode would accomodate a few bursts
over the Hubble time, and prevent too much chemical enrichment and gas consumption.

To investigate the possibility that gaps in the SFH are hidden in our data, 
we simulated a large number of synthetic CMDs.  
We used the simulator introduced in SHGC00, which evolved from the code
of Tosi et al. (1991) adapted to HST data by Greggio et al. (1998). 
In brief, the simulator is based on the evolutionary tracks of Fagotto et al. (1994),
and the tables of Origlia \& Leitherer (2000) for bolometric corrections.
The simulations
in this paper were carried out with an improved version of the simulator
described in SHGC00, leading to a better description of the
photometric errors. The error distributions and
completeness fractions in the simulator represent very well the I~Zw~36 data.

Being able to trace resolved stars with ages of at least 1~Gyr on the CMD,
we know that I~Zw~36 is not a ``primeval" galaxy. Lacking distinct main-sequence
turn-offs of an old stellar population (the data are not very deep as I~Zw~36 is
quite distant), we cannot tell the age of formation of the first stars.
In what follows, we therefore discuss cases in which the galaxy is modeled to
be ``young", as well as cases in which the galaxy is considered to be ``old".
We show what kind of models give good matches to the data, by comparing
the models to the data qualitatively, via CMD morphology, and quantitatively,
via luminosity functions (LF).

\subsubsubsection  {Young-galaxy models}

Several exemplary, synthetic CMDs are provided in Fig.~13.
For this particular set of examples, we used our low-metallicity (Z=0.0004)
simulator, with the appropriate, short distance modulus. We also adopted the
Salpeter initial mass function. Notice that
the simulator does not place TP-AGB stars on the CMD, as we do not
have stellar parameters for this phase. 
As in SHGC00, we keep track of potential
TP-AGB stars in a counter. We use the best available constraints on the masses
of stars which evolve into the TP-AGB phase (1.5 to 3~M$_\odot$), and the lifetime 
of the TP-AGB phase (1.5~Myr),
yet our simulations tend to always produce more synthetic stars in the TP-AGB phase
than are actually observed.

The simulations displayed in Fig.~13 consist of five examples.
Case (a) shows the CMD morphology for continuous star formation that started 2~Gyr ago. 
All other panels illustrate a short, 10-Myr long starburst, preceded
by a widening gap in star formation, preceded by the formation of the
remaining stars with ages up to 2~Gyr.  The gap lasts  
10~Myr for case (b), 100~Myr for (c), 500~Myr for (d), and
1~Gyr for (e). The star formation was completely stopped in the gaps, that is,
the SFR was dropped to 0~M$_\odot$~yr$^{-1}$ for the duration of the quiescent period. 

Notice how cases (a), (b), and (c) all produce acceptable CMD morphologies as compared
with the data. Cases (d) and (e), where star formation is interrupted for several hundred
Myr, exhibit a clear lack of RSG/AGB stars. In other
words, the red plume so well-populated in the data is cut away by
the cessation of star formation. This indicates that gaps of several hundred Myr
in recent history are disallowed by the data.

It is difficult to fine-tune the gap simulations and to derive more
precise gap durations, time placements, and amplitudes. For one, the time resolution of the 
data is not uniform across the CMD. The highest resolution occurs in most recent times. 
Another issue is the empirical adjustment of the SFR (and/or metallicity) 
on either side of the gap. Finally, dropping the SFR to 0~M$_\odot$~yr$^{-1}$ 
in the gap is an extreme assumption, which may not be realized in nature.
While we cannot explore the entire parameter space of possible gap durations, 
we can nontheless make some general statements as to what probably is, and what 
is not, compatible with the data.

All of the available data from the UV through the radio show evidence of massive stars.
Therefore, we cannot drop the SFR in the recent 10~Myr (or 100~Myr, etc.)
down to zero. Such a gap is not allowed. However, if we slide the 10~Myr gap
back by a mere 10~Myr in the simulator (case (b)), we already lose it in the data. As the time 
resolution decreases with age, we could also hide a 10~Myr gap on the CMD 
anywhere else at earlier times. Similarly, since we can hide a 100~Myr gap between 
10 and 110~Myr (case (c)), where time resolution is high, we can also introduce 
it at earlier times and it would go unnoticed in the data. We feel confident 
that we can rule out discontinuities in the SFR of the order of several 100~Myr, 
within at least the recent Gyr. In addition to the simulation shown as case (d), 
we experimented with a 500~Myr gap that we slid back in time within the recent Gyr. 
We find that a 500~Myr gap is too long to be hidden by the data. 
We conclude that somewhere between 100 and 500~Myr, we transition to a gap
length (for a zero SFR gap) that cannot be hidden anywhere in the recent Gyr.
Finally, it is also impossible to stop the star formation for more than a Gyr in the
recent 2~Gyr. If this gap occured in the most recent Gyr, it  would 
push about 90\% of the stellar content 
into the red tangle below the TRGB, something that clearly disagrees with the data (case (e)).
Even placed elsewhere, such a long gap clearly unbalances the morphological appearance of the
CMD. In general, the red tangle area of a CMD provides virtually no age resolution. It
can in principle contain stars with ages similar to the Hubble time. We can hide
Gyr-long gaps here, or drop the SFR to more realistic ``low states" for extended periods of
time, and we have no way of noticing.

We now provide some details on how we ran the simulations, and on the SFRs which
ensued. When we assume continuous star formation which started 2~Gyr ago (case (a)), then
the SFR implied by our CMD is typicially about 
4.9($\pm$0.3)$\times$10$^{-3}$~M$_\odot$~yr$^{-1}$. The error reflects the statistical
variations that come about when we run a large number of simulations with identical
parameters. The 2-Gyr-constant-SFR model is qualitatively a good match of the CMD, and 
quantitatively a good match to the LF. This is illustrated in Fig.~14, where we compare the
LF of the data with the LF of a 2-Gyr-continuous SFR model. The data and the model were binned
in 0.25~mag. intervals, and the errorbars shown are $\pm$ the squareroot of the
starcount in each bin. Notice in particular that it was possible 
to match the observed power at the TRGB (where we have a large starcount)
to within the respective errors. In a large set of simulations, 
some individual simulations showed deviations at the TRGB;
these were smaller than 20\% and in the direction of generating too many stars
just above the TRGB and too few just below the TRGB.

Between about 50--100\% of the stars in the blue 
plume could potentially be massive stars. If all of the stars in the blue plume are indeed
massive MS stars and BSGs with ages no larger than 10~Myr, then there is a high power 
in the ongoing event.
To simulate and test the burst+quiescent scenario, we indeed put all of the power 
consistent with the number of stars in the blue plume (i.e., 10\% of all stars on the CMD) into
the young burst. (An alternative, discussed below, is to assume that the
ongoing event started earlier than 10~Myr ago, and that some of the stars at the bottom of the
blue plume should be interpreted as BL stars.) We adopted 10~Myr for the
duration of the event, since this is typical of the length associated with
the starburst in I~Zw~36 by Fanelli, O'Connell \& Thuan (1988),
$\leq$10~Myr, and Deharveng et al. (1994), $\leq$12~Myr. With these
assumptions, the present-day star formation rate indicated by the blue plume of
the NIC2 CMD is 2.5($\pm$0.4)$\times$10$^{-2}$~M$_\odot$~yr$^{-1}$, or about a factor
of 5 up from continuous star formation over the last 2~Gyr. 
The error on the SFR again reflects stochastic variations, and was derived
by running a large number of simulations with the above parameters. 
For each of the cases (b) through (e) shown in Fig.~13, we used a ``fresh" 
simulation for the 10-Myr starburst, 
allowing us to illustrate the statistical
effects in the supergiant population (cf. Greggio 1986). The SFR
given above is the highest rate that is consistent with the number of observed stars.
Cases using the maximum SFR can produce satisfying synthetic CMDs and LFs. For example,
we achieved an excellent match to the data with a model that has a high-power starburst
preceded by a 100~Myr gap. In this model, the SFR in the 10-Myr burst was  
2.1$\times$10$^{-2}$~M$_\odot$~yr$^{-1}$, and the SFR from 110~Myr to 2~Gyr 
was 4.4$\times$10$^{-3}$~M$_\odot$~yr$^{-1}$. This implies that solutions in which
the ongoing event is very short and by about a factor of 5 stronger than the average recent
SFR are allowed.

If we assume instead that fewer than 100\% of stars in the blue plume 
are in reality massive,
young stars born in the recent 10~Myr, and allow a longer duration for the
ongoing event, then the SFR goes down. The duration of the ongoing
event is not constrained by our data. Recall that a 2-Gyr-continuous SFR model
matches the CMD and the LF of the data. If the star formation extended
to earlier times, then
some of the stars at the bottom of the blue plume would be
interpreted as older, BL stars at the blue edges of their blue loops
(cf. Fig.~12 of SHGC00). 
Since there are not a lot of stars in the blue plume in total,
we can cut the SFR derived for the 10-Myr-burst scenario and still 
easily make up for the few missing 
faint stars with a contribution by BL stars. 

To summarize, the near-IR CMD does not constrain well a recent, short
star-forming event. We can accomodate either a long episode at a relatively low
rate, or allow for a very recent SFR enhancement, of virtually any factor
up to 5. The maximum current SFR compatible with our CMD is the same
as the one derived from UV and radio data. 

\subsubsubsection  {Old-galaxy models}

So far, we have treated I~Zw~36 as a young galaxy with no star formation
earlier than 2~Gyr ago. We emphasize that a young model galaxy with a more 
or less continuous SFR can provide a good match of the data. The young model is our
conservative model, since we cannot really proove the existence of
ancient stars with the CMDs in hand. 

However, we have circumstantial evidence
which indicates I~Zw~36 may be an old galaxy. We 
know that stellar evolutionary tracks for low masses and hence, old
ages, do pass through the red tangle of the CMD. We also know that the integrated
color of the background sheet at large distances from the star-forming core
can be identified with an old stellar component. 
Therefore, we also ran several simulations which extended the 
SFH further back than 2~Gyr (to 5, 7, and 15~Gyr). 

These simulations are constrained by the fact that we always need to
produce 712 survivors on the CMD. Therefore, in general, the older the 
stellar population, the lower the SFR.
We show in Fig.~14 the LF of a simulation for a 15-Gyr-continuous SFR model. 
We see that this simulation has too much power at the
TRGB, and produces too few stars of intermediate and high luminosity
above the TRGB. This implies that if we assume that I~Zw~36
is an old galaxy, then some amplitude variations in SFR
are required on long time scales. Specifically, for ages larger
than 2~Gyr, we need more power in the
recent history, and less power in ancient history, than what is provided
by models with continuous SFRs.

We therefore explored simulations in which we assumed that the
galaxy is old, and in which we adjusted the recent and early SFRs separately
to match the CMD and the LF.
We define as the recent SFR the SFR between 0 and 1~Gyr, and as the early SFR the
SFR from 1 to 15~Gyr. This allows us to derive constraints on the ratio
of SFR/$<$SFR$>$$_{past}$,
the birthrate parameter, directly from the resolved stellar content.
 
A natural dividing line between SFR and $<$SFR$>$$_{past}$ occurs indeed at an age of 1~Gyr
since at this age, RGB stars start to populate CMDs at the TRGB in large numbers.
We used the technique of the SG and AGB boxes developed in SHGC00. We assigned all stars
counted within these boxes to the recent Gyr. Specifically, this part of the model
has a constant star formation which started 1~Gyr ago and is constrained by
having to yield 20\% of its survivors with magnitudes H$_0$$<$23.3 and colors
-0.5$<$(J-H)$_0$$<$1.0. When simulations are run, they bring
along also the progenitors of theses stars which are below the TRGB. We then 
assumed that the remaining stars below the TRGB not yet accounted for have ages 
between 1 and 15~Gyr, and simulated them in the second part of the modelling
process. In this way, we are able to estimate the
contrast between SFR and $<$SFR$>$$_{past}$ from the CMD with a reasonable set
of assumptions. (Refer to SHGC00 for examples of how to hide variations in the SFR among the
early SFH of a BCD, also for comments on how these affect the total astrated mass
of a galaxy.) 

The LF of a model from our run of 14+1-Gyr-continuous scenarios
is shown in Fig.~14. We find that such models can provide 
good matches of the CMD and the LF of the data, with a quality that equals
the 2-Gyr-continuous models. These simulations yield a recent SFR which is 
5.7($\pm$0.8)$\times$10$^{-3}$~M$_\odot$~yr$^{-1}$;
and an early $<$SFR$>$$_{past}$ which is 
4.3($\pm$0.6)$\times$10$^{-4}$~M$_\odot$~yr$^{-1}$.
The ratio SFR/$<$SFR$>$$_{past}$ 
is about a factor of 13. The estimated factor of 13 for I~Zw~36 is very similar to our
result for Mrk~178 (SHGC00), where SFR/$<$SFR$>$$_{past}$ is 8---12 depending on the assumed
metallicity. 

To summarize, models which match the data can be obtained by extending
the SFH to old ages, as long as the power in the young and old stellar content are adjusted
appropriately. Being sensitive to SFR in the recent 1~Gyr, these CMD simulations
are constrained by the power consistent with star formation in the
recent Gyr, and leave the remaining stars to be assigned to any potential
old SFH. In general, SFR/$<$SFR$>$$_{past}$ increases  from 1 when the initial time 
of star formation is 2~Gyr, to an order of magnitude when the initial time is 15 Gyr. 
The trend basically reflects
the increase in the temporal range over which the star formation is demanded to
produce the observed number of stars below the TRGB. Both kinds of scenarios,
young and old, match the CMD and LF of the resolved stars. 
The morphology and integrated colors of
the background sheet lead us to prefer an old galaxy over a young galaxy.

\subsubsubsection  {Modelling summary}

This case study of I~Zw~36 allows us to gain insight into the star formation mode
of BCDs. Because I~Zw~36 is the most distant BCD in our NICMOS survey, its data
have the largest errors and incompleteness fractions. 
Gaps would be easier to hide in this data set, than in the CMDs of 
VII~Zw~403 and Mrk~178. As explained above, I~Zw~18 is at such a much larger distance
that its near-IR CMD is not directly comparable with the I~Zw~36 data or
simulations.
  
A model which assumes a single, short burst preceded
by a long quiescent period does not match the observations. Our conclusions are 
consistent with those reached by Aloisi, Tosi \& Greggio (1999) for the SFH 
of I~Zw~18 based on models of its optical CMD. 
This causes somewhat of a paradigm shift in the way we view BCDs.
Scenarios which consider current BCD activity as several-orders-of-magnitude 
strong ``delta functions" in the SFR do not give an accurate record of what we see on 
CMDs. 

Good models of the observed CMDs can be produced with more-or-less continuous SFRs
in recent times.  The recent mode of star formation of BCDs 
was therefore either continuous with modulations by factors of a few (thus allowing for some
``burstiness"), and/or ``gasping" in the terminology of Tosi (1998). In particular, however,
in $\it{all}$ of the BCDs resolved with HST, star formation is seen to have been active for
at least 1~Gyr. BCDs are unlikely to be truly primeval galaxies, although
they may be fairly young.

We consider the present activity part of an
extended high state. Extended periods of low states in SFR may have occured
at early times, if BCDs are old galaxies, but deeper single-star photometry is 
needed before we can derive the age of formation of BCDs. 
The CMDs allow us to nevertheless constrain the average powers in the high and low
states. For instance, we may assume that BCDs are ancient galaxies,
having formed their first stars as much as 15~Gyr ago. If this is the case, then
we derive a factor of about 10 for SFR/$<$SFR$>$$_{past}$ for I~Zw~36, similar
to what we inferred in SHGC00 for Mrk~178.

\section{Discussion}

I~Zw~36 is observed with HST to deep enough limiting magnitudes to resolve 
stars in the RGB phase. 
We consider it our most important finding that whenever deep enough observations
are available, we discover RGB stars in BCDs. These stars bear
witness to star formation which predates the ongoing activity. The RGB stars can in
principle be much older than the minimum age of 1-2~Gyr that we are
able to assign to them based on our CMDs. However, we require even deeper CMDs able 
of resolving stars on the blue horizontal branch or an old main-sequence turn-off, 
before we can ascertain the presence of truly ancient ($>$10~Gyr) stars. Such data are beyond the 
observing capabilities of the present time. Nevertheless, we think that the combination
of resolved RGB stars together with a low-surface-brightness, red halo, makes
a strong case for an underlying population of old stars in I~Zw~36.

Three BCDs have now been studied well below the TRGB with NICMOS: VII~Zw~403, Mrk~178, and I~Zw~36.
They range in metallicity (O abundance of the ionized gas) from Z$_{\odot}$/10
to Z$_{\odot}$/20. They are on the low-metallicity side of the observed peak in
O-abundance distribution of BCDs (e.g., Izotov \& Thuan 1999). It is this low O abundance
which has traditionally been the driver for the youth hypothesis.
In the alternative hypothesis, assuming BCDs are old galaxies, they should have had only a few,
short starbursts separated by long quiescence periods, else they enrich
too quickly and/or run out of gas (see, e.g., the review of Thuan 1991).
The old galaxy hypothesis is favored by our finding of a red backgound
sheet which extends far beyond the core of star-forming activity.

What then, are the strengths and duty-cycles of BCD starbursts?
BCDs are often thought of as intensely star-forming systems. However,
the recent star formation rates of VII~Zw~403, Mrk~178, I~Zw~36, and I~Zw~18
are only several 10$^{-3}$ to a few 
10$^{-2}$~M$_\odot$~yr$^{-1}$. 
The star formation rates of these BCDs are quite moderate. Given moderate star formation
rates, these BCDs should not have any difficulty economizing their gas 
supply for more than another to up to several Gyr. 

More specifically, the model SFRs derived for I~Zw~36 shed 
some light on its gas-consumption time.
At a high SFR, 2.5$\times$10$^{-2}$~M$_\odot$~yr$^{-1}$, consistent with
a 10-Myr burst, and assuming 100\%
efficiency, the available 6x10$^7$M$_{\odot}$ of H-I gas will be consumed
in about 2.5~Gyr. However, it is clear that such a burst has a short duration.
The average SFR over the most recent Gyr is 
5.7$\times$10$^{-3}$~M$_\odot$~yr$^{-1}$. If I~Zw~36 continues
to make stars at this average rate, its H-I supply will last 
for another 10~Gyr.  In other words, excessively rapid gas consumption
does not have to be taking place, and thus does not force us to adopt episodic
star formation with short bursts and long quiescent periods as the star 
formation mode of BCDs (although one can envision
that all gas available for star formation will eventually be used up).
Finally, if we assume a large formation age, e.g., of 15~Gyr, then the $<$SFR$>$$_{past}$ 
consistent with the red tangle
area of the CMD is an order of magnitude below that of the current high state,
implying a gas-consumption time much longer than a Hubble time. Thus an
earlier, low state of star formation insures that BCDs are gas rich (and 
metal-poor) in the present epoch.

Simulations of the CMDs of VII~Zw~403, Mrk~178, I~Zw~36 and I~Zw~18  have shown that
star formation is consistent with a more or less continuous history at least over the past Gyr. 
This is what we call the high state.
Fig.~13 illustrates how continuous star formation is that mode which best 
represents the near-IR CMD morphology for I~Zw~36, although interruptions in star 
formation can be supported by the data, as can be short bursts if they
have less than about an order of magnitude in power compared to the recent average rate. 
Whereas BCDs as a class have been
viewed as systems undergoing short and strong starbursts separated
by long quiescent periods (e.g., Searle, Sargent \& Bagnuolo 1973; Thuan 1991; 
Mas-Hesse \& Kunth 1999), our 
observations indicate that such behavior may not be the norm.

The BCDs I~Zw~36, Mrk~178, and VII~Zw~403
belong to the iE type, which comprises 70\% of all BCDs.
If they are representative of the iE types as a whole, then it 
stands to reason that indeed the majority of BCDs do not undergo short bursts of star 
formation separated by Gyr-long periods of quiescence, as has been our common conception 
of the star formation histories of BCDs. I~Zw~18 belongs to the rare few percent of
BCDs which are still considered to be candidate primeval galaxies, although there is
much controversy on its age of formation. While its optical and near-IR 
CMDs clearly indicate that star formation began at least a Gyr ago, if not earlier, 
the results for the mode of star formation are not as yet conclusive. 

We cannot comment on the SFH past the ``1-2~Gyr frontier" of the
CMDs for any BCD. However, for I~Zw~36 and Mrk~178, we provide an estimate
of SFR/$<$SFR$>$$_{past}$ based on resolved stellar content. If we assume
that these galaxies are as old as 15~Gyr, then the contrast between the present
high state and the past low state is as high as about an order of magnitude.
The possibility of extended star formation
will require us to look for ways in which we
can reconcile the low O abundances seen in the ionized interstellar gas of BCDs with
a more or less continuous history of star formation. This may be possible if
star formation rates in early and extended low states are small enough
(Legrand et al. 2000; Legrand 2000; Henry, Edmunds \& K\"{o}ppen 2000).  
Indeed our constraints for I~Zw~36 and Mrk~178 give low states (for an age
of 15~Gyr) of a few times 10$^{-4}$~M$_\odot$~yr$^{-1}$, similar to what 
Legrand's model assumes for the early SFH of I~Zw~18. 

The nature of our data does not allow us to provide insight on details of the SFR 
early in the history of BCDs (see SFGC00). In order
to achieve that, CMDs with much deeper limiting magnitudes are needed.
We note, however, that even among the 20 or so dwarfs of the Local Group
studied to deep limiting magnitudes (see the recent
review of Grebel 2000), an episodic
mode of star formation is the exception rather than the rule. Among the 
Local Group dwarfs, the Carina dSph is presently
the only convincing case of a dwarf galaxy which experienced several 
distinct episodes of star formation 
separated by Gyr-long
quiescent periods (Smecker-Hane et al. 1994; Hurley-Keller, Mateo \& Nemec 1998). 
Since episodic star formation is so rare in Local Group 
dwarf galaxies, be they dIrr, dwarf Spheroidal (dSph), or dIrr/dSph in type, it is perhaps not 
surprising that it appears to be 
uncommon among the BCDs.

We have been interested in investigating the extent to which the SFHs of BCDs 
are compatible with models of faint blue excess galaxies. For instance, 
BCDs could provide alternative remnants for the delayed-formation scenario of dwarfs.
This scenario attributes faint blue excess
galaxies to a population of bursting, blue dwarfs at intermediate redshift which
has, by and large, faded and/or merged by the present epoch (e.g. Babul \& Ferguson 1996). 
Today's dSph galaxies have been considered front runners for the 
faded, red remnants (e.g. Guzm\'{a}n et al. 1998). 
However, galaxy counts in the Hubble Deep Field (HDF) have demonstrated that
models in which star formation takes place in a single, short burst followed
by fading, overproduces faint red remnants with respect to the observed counts 
(Ferguson \& Babul 1998). We have considered that BCDs could
provide some of the remaining, and recently ``rejuventated", 
remnants of faint blue galaxies. Greggio et al. (1998)
investigated the dIrr galaxy NGC~1569, and concluded that its stellar content is
consistent with the kind of starburst required in the Babul \& Ferguson model.
Aloisi, Tosi \& Greggio find that the SFR of I~Zw~18 falls short by 
2 orders of magnitude to make it a local counterpart of the faint blue galaxies 
according to the Babul \& Ferguson model.
We tested the Babul \& Ferguson
scenario on the CMD of Mrk~178, and found that it was difficult but not impossible to achieve
a burst of the necessary strength among the RGB population (SHGC00). The general
lack of a highly episodic nature of the star-formation rates of local
dwarf galaxies, including BCDs, would seem to speak against the scenario
in which the faint blue galaxies are briefly bursting dwarfs.

The evidence which accumulated over the past few years now suggests
that neither whole-sale merging of dwarfs (Rocca-Volmerange \& Guideroni 1990),
nor strong luminosity evolution in dwarfs (Phillips \& Driver 1995), nor a population of
blown-away dSphs (Babul \& Rees 1992), can account for all aspects of the faint blue galaxy excess.
If BCDs are predominantly 
galaxies with 
gentle star formation variations, then a steeper faint end of the local, 
optical galaxy luminosity function (Gronwall \& Koo 1995) may provide the most
natural explanation of the faint blue excess galaxies. 

Extended periods of activity would also help to explain why three decades of work have not
revealed any obvious candidates for BCDs in quiescence in the local Universe 
(Kunth \& \"{O}stlin 2000).

\section {Conclusions}

I~Zw~36 is the third low-metallicity BCD of the iE subtype in which the RGB is revealed with
HST/NICMOS. The data indicate stars with ages of at least 1-2~Gyr, and a few
lines of evidence hint at an even earlier age of formation.
The ongoing star formation in the galaxy core, including Wolf-Rayet stars, 
occurs within an extended body of older stars. These low-mass stars make the dominant
contribution to the stellar mass of the galaxy.

We emphasize the usefulness of near-IR CMDs in detecting the evolved 
descendants of intermediate and low-mass stars whose spectral energy 
distributions peak in the near-IR. Thus, near-IR CMDs
complement near-UV CMDs, which preferentially sample the young populations
of hot, massive stars; they help to avoid the problems of 
color biasing.

While I~Zw~36 is actively forming stars in its core, the present-day star formation
rate is not very high. At its
present rate of 2.5$\times$10$^{-2}$~M$_\odot$~yr$^{-1}$, and assuming a 100\%
efficiency, I~Zw~36 will run through its H-I supply of 6x10$^7$M$_{\odot}$
in about 2.5~Gyr. The limited gas-consumption time has been
one of the reasons why star formation in BCDs is considered to
proceed in a series of a few short bursts separated by long quiescent periods.
An alternative to this extreme scenario are more extended high and low states of
star formation. The average SFR which is consistent with the high activity over the
recent Gyr, for example, can continue for another 10~Gyr before it consumes all
of the H-I gas.

We address the mode of star formation of BCDs with simulated CMDs. As a case study, we
provide simulated star-formation histories of I~Zw~36. We show that if we
allow short stops (10 --- 100~Myr) in its recent star formation history
we can still produce CMDs which match the data. However, once star formation
is interrupted by several 100~Myr, the morphology of the CMD changes so
drastically that we conclude long gaps with zero star formation cannot be hidden in the data. 
We find for all three iE BCDs investigated by our team not only that the present star 
formation does not represent the first event in their history, but also 
that star formation has not proceeded, at least in recent times, in a 
series of short, strong bursts with long quiescent periods.
A moderate modulation in the recent SFR is consistent
with our data. We also gleaned some information about possible,
ancient star formation from the red giants of I~Zw~36. Assuming an age of 15~Gyr, we divided
the SFH into the interval from 0-1~Gyr, and from 1-15~Gyr; and
we derived SFR/$<$SFR$>$$_{past}$ = 13. If I~Zw~36 is indeed an ancient galaxy,
then the SFR in the recent Gyr implied by its CMD is by as much as an order of magnitude 
higher than its average past SFR, in the region encompassed by the field of view
of the NIC2 detector. We previously determined a similar birthrate 
for Mrk~178.

Extended periods of high- and low-states of star formation, which we infer from model
CMDs to be viable SFHs, would alleviate the
problem of the missing quiescent counterparts of BCDs in
the local Universe, and have implications concerning models in which faint blue excess
galaxies are bursting dwarfs. They also challenge star formation histories
of BCDs based on chemical-evolution modelling, by allowing for the existence
of gas-rich and metal-poor galaxies which are nevertheless not young. 

\acknowledgments We thank Dr. Deidre Hunter for sharing the H$_{\alpha}$ 
flux of I~Zw~36 prior to publication. We also thank Dr. Andrew Hopkins
for help in extracting and interpreting the radio data. 
Work on this project was supported 
through HST grants to RSL and MMC (projects 7859 and 8012). 
UH acknowledges financial support from SFB~375. 
LG acknowledges support from
the Alexander von Humboldt Stiftung.
We made extensive use of the SIMBAD and NED data bases.

\clearpage

\figcaption[] {The 103a-E image of I~Zw~36 from the Digitized Sky Survey
in grey-scale representation. The brightest feature is a blend of several
H-II complexes. This is where the NIC2 (red) and FOC (blue) cameras were pointed. 
The faint, elliptical main body can also be seen. The NIC1 (small red box)
was pointed in the background sheet. 
}

\figcaption[]{a) NIC2 image composed by combining all images taken
through the F110W filter. b) Same for images taken through the
F160W filter. The galaxy is clearly resolved into stars.}

\figcaption[]{The DAOPHOT-derived photometric errors.}

\figcaption[]{Completeness fractions from ADDSTAR tests.}

\figcaption[]{Color-magnitude diagrams of I~Zw~36 in magnitudes
in the HST Vegamag system. The location of the TRGB is indicated.}

\figcaption[]{Color-magnitude diagrams of I~Zw~36 after transformation
to ground-based J and H. The foreground extinction is negligible. }

\figcaption[]{Luminosity functions of I~Zw~36 in instrumental HST magnitudes
in the Vegamag system. 
The location of the adopted TRGB is marked. }

\figcaption[]{Color-magnitude diagrams of I~Zw~36 with three cuts applied on the
size of the photometric errors (1${\sigma}$) in both J and H.}

\figcaption[]{Color-magnitude diagrams of I~Zw~36 with two
sets of tracks overlayed. The first set of tracks is for a metallicity
of Z=0.0004 and the corresponding short distance modulus of 28.8. The
second set is for Z=0.004 and the long distance modulus of 29.5. 
For each case, the TRGB is indicated, and the corresponding 
absolute-magnitude scale is given on the right-hand ordinate. }

\figcaption[]{NIC2 camera X- and Y-positions in the blue plume
(plotted here are stars with (J-H)$_0$$<$0.20) and in the red tangle
((J-H)$_0$$>$0.5 and H$_0$ larger than the TRGB). The young, massive stars in
the blue plume occupy the central part of the image, in the same region
where the FOC images also show massive stars and ionized gas (compare Figs. 1 and 2).
The stars in the red tangle fill the image almost uniformly (with the exception 
of those central portions occupied by H-II regions
where there is more crowding). This is the morphological feature
that we refer to as Baade's red sheet. }

\figcaption[]{The near-UV---optical, FOC CMD of I~Zw~36 from Deharveng et al. (1994) is compared with
the near-IR, NIC2 CMD. The CMDs are plotted in such a way that their blue
and red plumes roughly overlay. Notice the blue plume dominates the
near-UV---optical CMD, while the red plume is the most prominent feature of the near-IR CMD.
This may result in color biasing of interpretations of CMDs.}

\figcaption[]{A comparison of the near-IR CMDs of VII~Zw~403, Mrk~178, and I~Zw~36. 
For more details, refer to the text. The TRGB is indicated
by a dashed green line. The dashed blue line indicates the location of the
blue plume. The black dashed line is our dividing line
between RSG (left) and AGB (right) stars.}

\figcaption[]{Exemplary, simulated CMDs for I~Zw~36, using our Z=0.0004 simulator.
Case (a) -- continuous star formation from 2~Gyr ago until today. Cases (b)--(e), 
quiescent periods of 10, 100, 500, and 1000~Myr
prior to the present star formation. The CMD morphology of cases (d) and (e)
clearly differs from that of the data, effectively ruling out any long discontinuities
in the recent star formation history of I~Zw~36.}

\figcaption[]{We judge the quality of a model by how well it reproduces the 
CMD morphology and the luminosity
function of the data. The LF of the data is shown by the heavy black line. Starcounts
are given in 0.25 mag. intervals; the errorbars drawn are $\pm$ the squareroot
of the starcount in each bin. Several models are overplotted on the data. Notice how
the model with a 2-Gyr-continuous SFR provides a good overall match to the data.
We can interpret this galaxy as a ``young" galaxy. Several lines of evidence
indicate the interpretation that the galaxy is ``ancient" is also possible. 
However, a model with a 15-Gyr-continuous SFR does not match
the data; there is too much power in the old stellar component at the TRGB and too 
little power in the young and luminous stellar content. 
On the other hand, a good match is achieved when the powers are adjusted such that
SFR/$<$SFR$>$$_{past}$=13. 
This is illustrated by the 14+1-Gyr-continuous SFR model.}

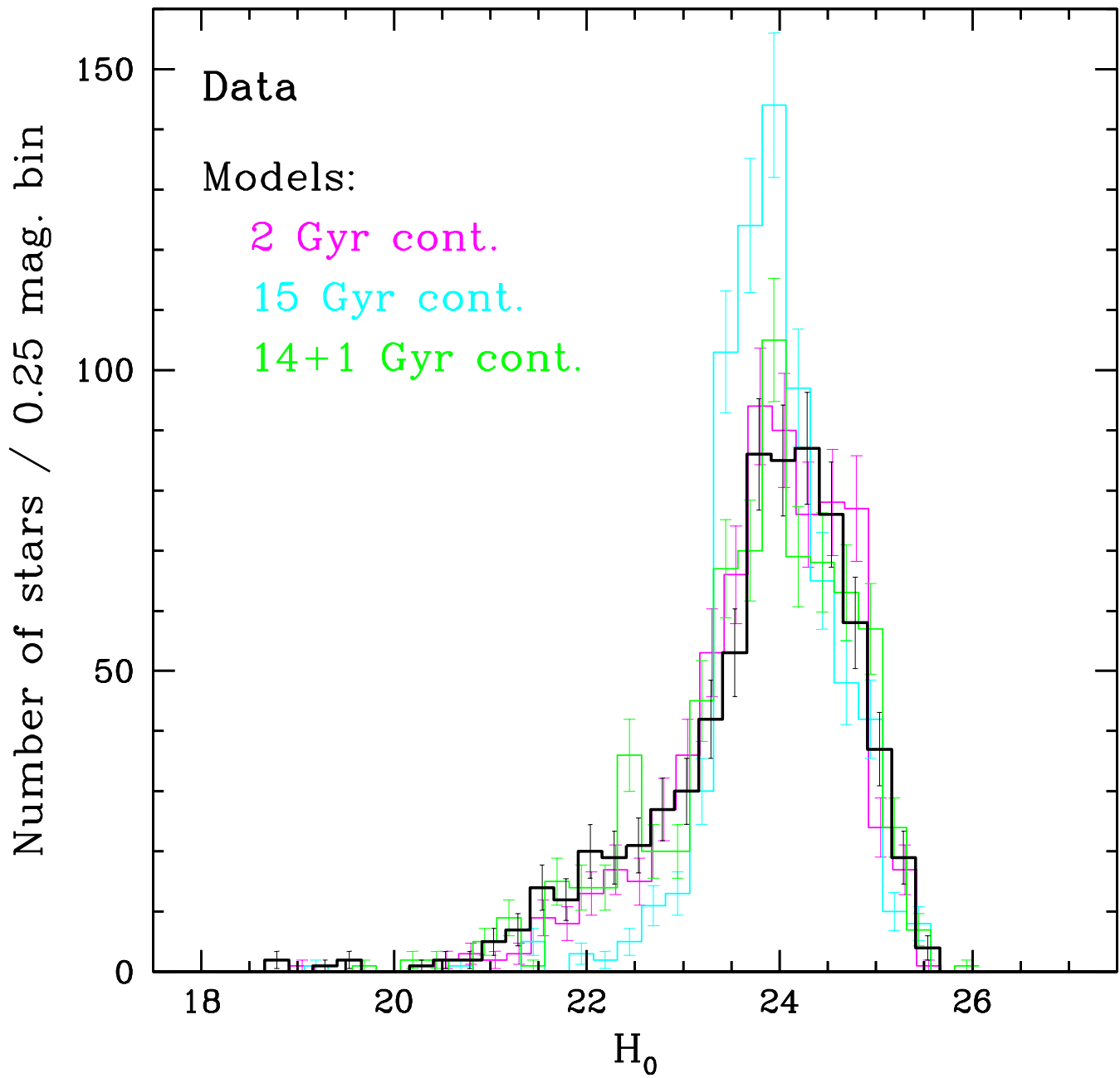
\end{document}